\documentclass{article}  
\usepackage{jamaica04}

\newcommand{\xfig}[1]{\begin{center}
\mbox{\epsffile{#1}}
\end{center}}
\def\bea{\begin{eqnarray}}
\def\eea{\end{eqnarray}}

\frompage{000} \topage{000}                                              
\def\rightmark{Event Structure at RHIC from p-p to Au-Au}
\def\leftmark{T.\,A.~Trainor}
 
\title{Event Structure at RHIC from p-p to Au-Au} 
\authors{
{Thomas A. Trainor$^1$ (STAR Collaboration)
}\\[2.812mm]
{\normalsize
\hspace*{-8pt}$^1$CENPA 354290, University of Washington, \\ 
 Seattle, USA\\[0.2ex] 
}}
 
\abstract{Several correlation analysis techniques are applied to p-p and Au-Au collisions at RHIC. Strong large-momentum-scale correlations are observed which can be related to local charge and momentum conservation during hadronization and to minijet (minimum-bias parton fragment) correlations.}

 
\begin{document}
 
\maketitle
\setcounter{page}{1}

\section{Introduction}\label{intro}

Two-particle momentum distributions from p-p and Au-Au collisions are projected onto subspaces $(y_t \otimes y_t)$ and $(\eta_\Delta\otimes \phi_\Delta)$ for charge-independendent (CI or isoscalar) and charge-dependent (CD or isovector) charge combinations, revealing strong correlations related to parton dynamics, a color medium and changes in hadronization.
 
\section{p-p Collisions: The Essential A-A Reference}

\begin{figure}[h]
\vspace{-.1in}
\begin{tabular}{ccc}
\begin{minipage}{.31\linewidth}
\epsfysize 1\textwidth
\epsffile{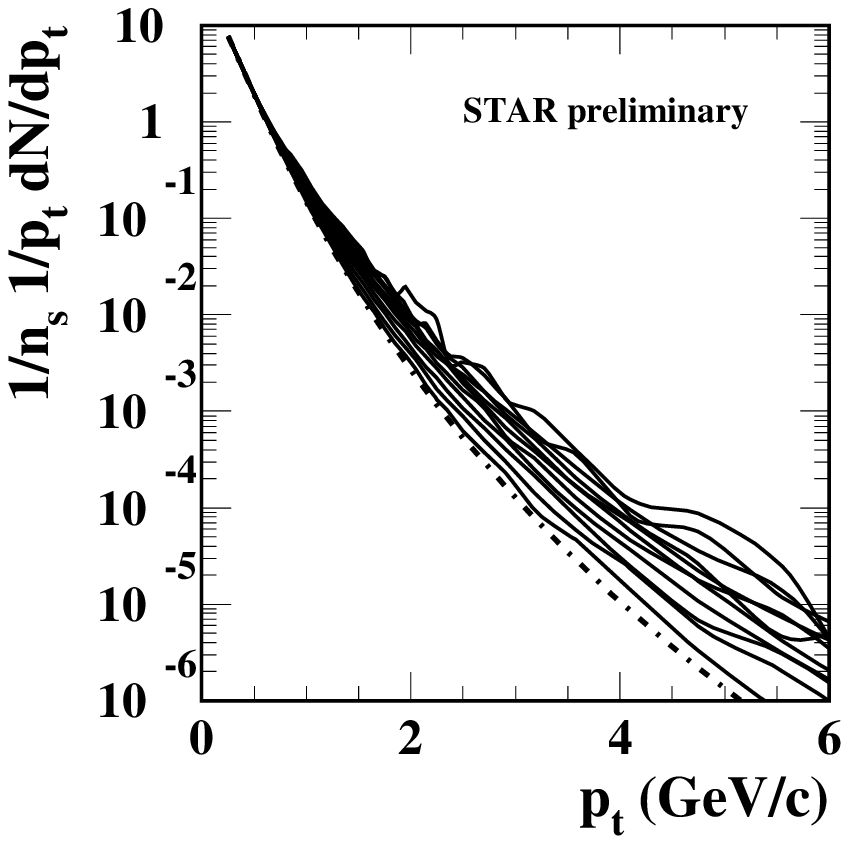}
\end{minipage} &
\begin{minipage}{.31\linewidth}
\epsfysize 1\textwidth
\epsffile{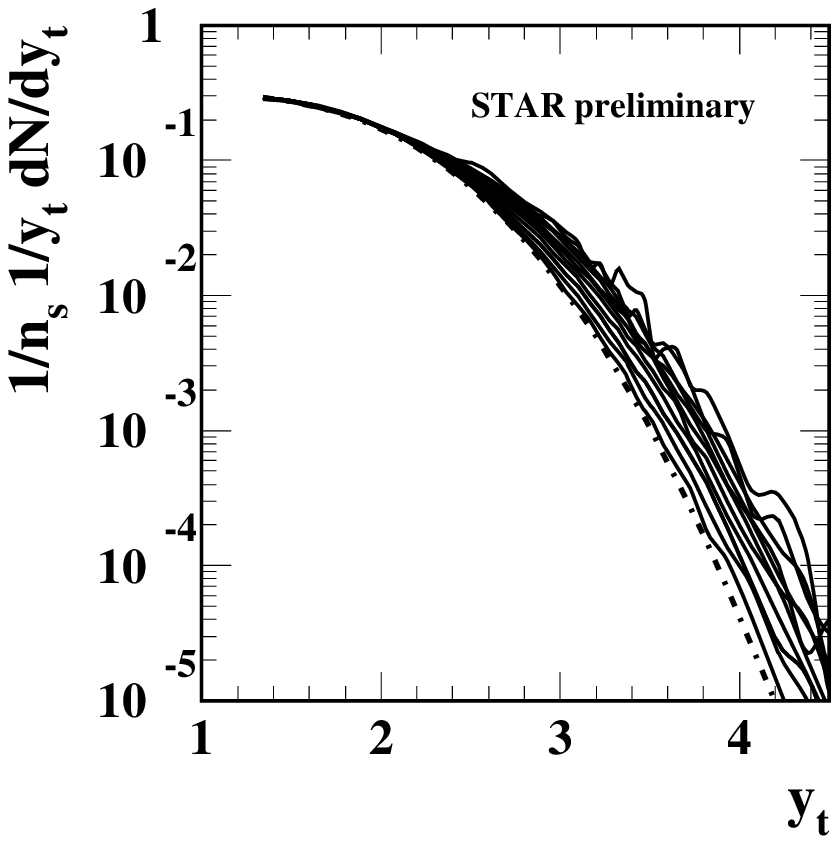}
\end{minipage}
\begin{minipage}{.31\linewidth}
\epsfysize 1\textwidth
\epsffile{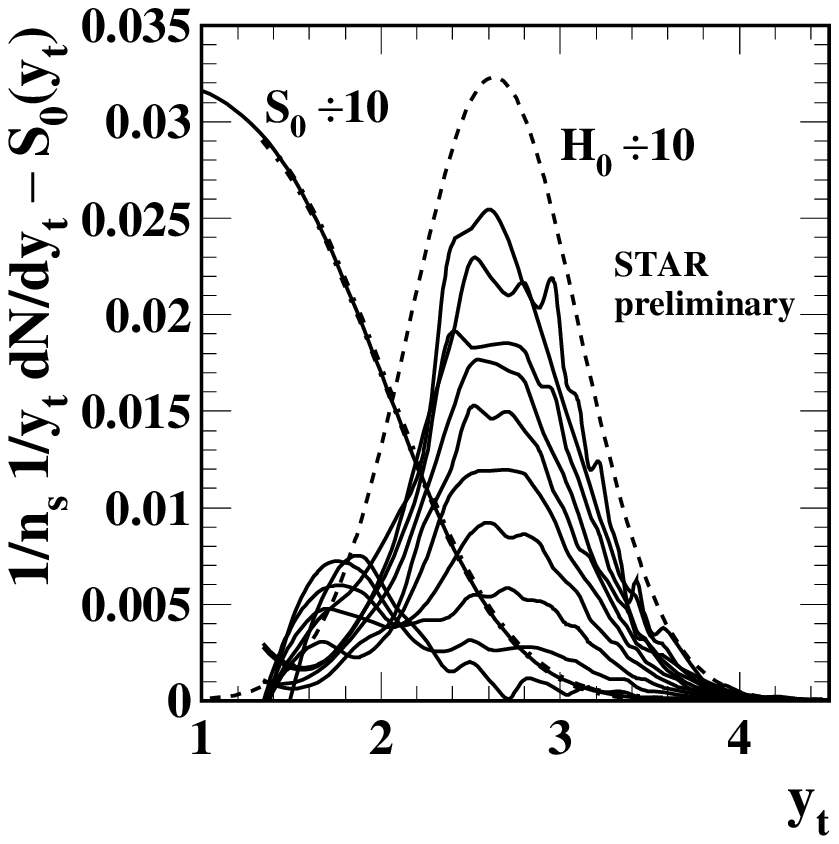}
\end{minipage}
\end{tabular}
\caption{Panels from left to right: inclusive $p_t$ distributions for $n_{ch} \in [1,12]$, same on transverse rapidity $y_t$, soft reference $S_0$ subtracted to reveal hard components.\label{tat-fig1}}
\end{figure}

The structure of hadron distributions from p-p collisions provides an essential reference for A-A collisions. Normalized $p_t$ distributions for several event multiplicity classes of p-p collisions from RHIC at $\sqrt{s} = 200$~GeV~\cite{pppt} shown in Fig.~\ref{tat-fig1} (left panel) can be decomposed into a soft component and a semi-hard or minijet component. p-p event multiplicity at this energy determines the frequency of hard scatters in each event class. A two-component model function is used to describe the data: $1/p_t\, dN / dp_t = n_s(n_{ch})\, S_0(p_t) + n_h(n_{ch})\, H_0(p_t) $, with soft and hard reference components $S_0(p_t)$ and $H_0(p_t)$ (independent of $n_{ch}$ by hypothesis) both integrating to unity on $p_t$. Transforming to transverse rapidity $y_t \equiv \ln\{(m_t+p_t) / m_0\}$ as a velocity variable provides a `native' description of minijets as hadron fragments from a moving source. We observe in Fig.~\ref{tat-fig1} that soft-component $S_0$ (dash-dot curve, string fragments) is well represented on transverse rapidity $y_t$ by an error function (corresponding to a L\'evy or `power-law' distribution on $m_t$), and hard component $H_0$ (dashed curve, minijet fragments) is represented by a gaussian distribution on $y_t$, each form approximately independent of event multiplicity and determined by two parameters. The stability of the minijet fragment distribution with event multiplicity (right panel = center-panel $-$ $S_0(y_t)$) shows that gaussian distribution $H_0$ on $y_t$ is a good minimum-bias minijet input for A-A collision models.

\begin{figure}[h]
\vspace{-.0in}
\begin{tabular}{ccc}
\begin{minipage}{.3\linewidth}
\epsfysize 1\textwidth
\xfig{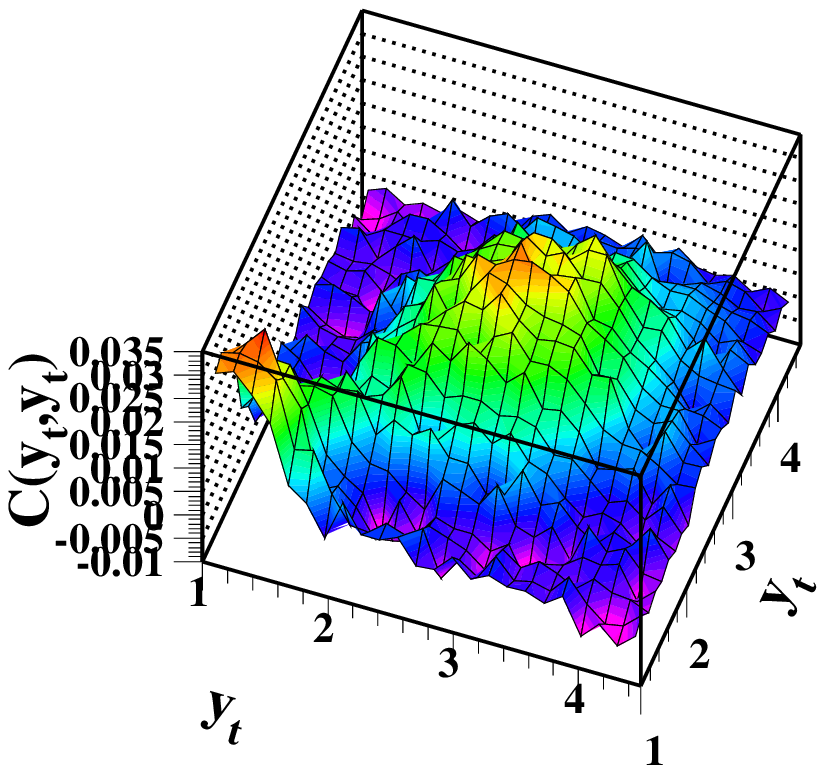}
\end{minipage} &
\begin{minipage}{.3\linewidth}
\epsfysize 1\textwidth
\xfig{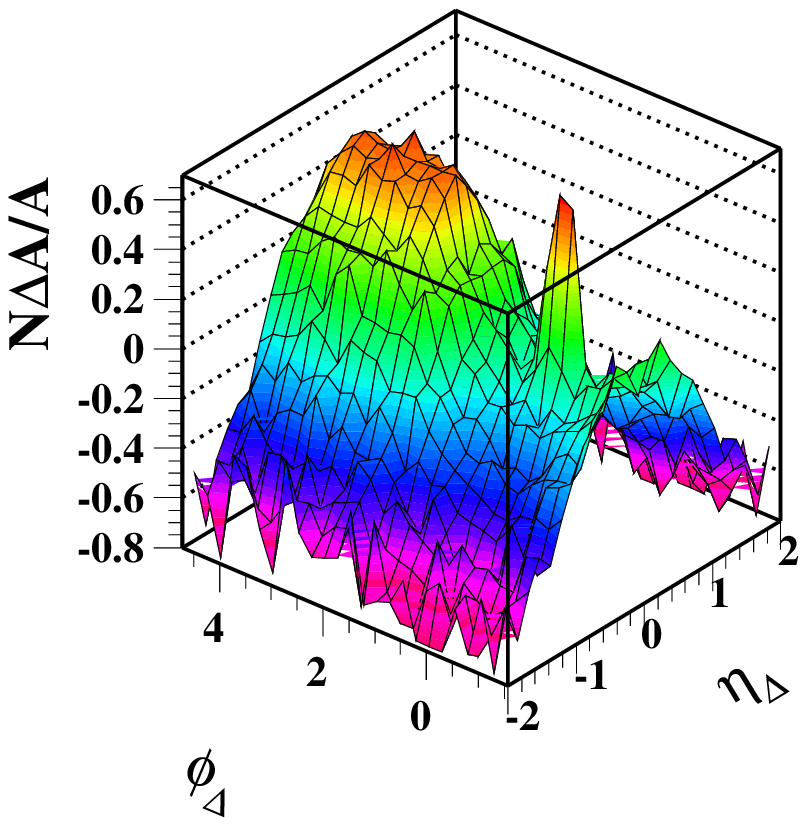}
\end{minipage} &
\begin{minipage}{.3\linewidth}
\epsfysize 1\textwidth
\xfig{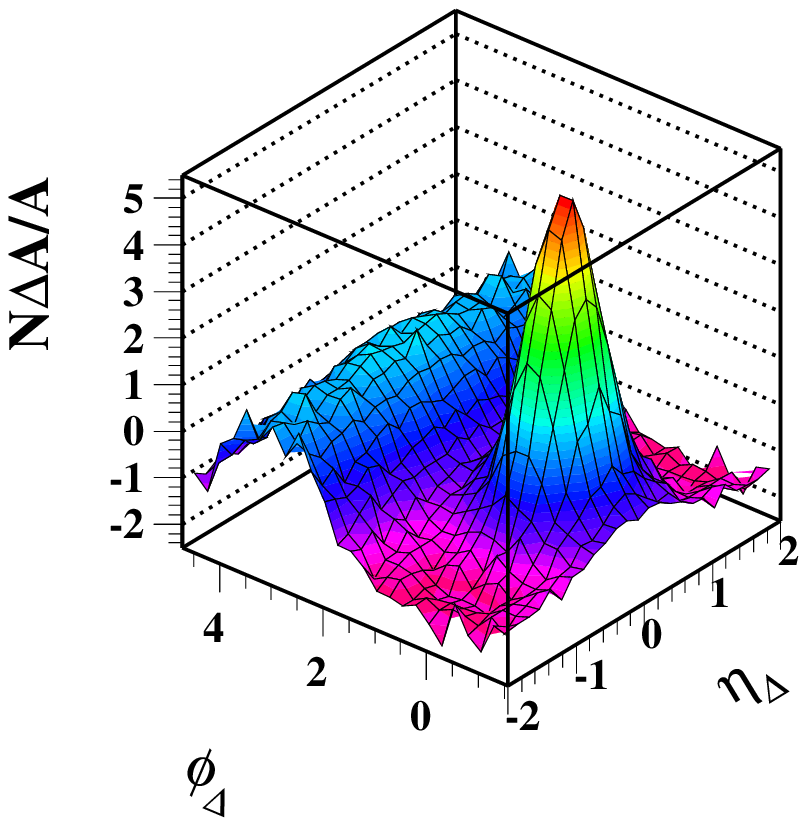}
\end{minipage}
\end{tabular}
\caption{$y_t \otimes y_t$ correlations (left) and $\eta_\Delta \otimes \phi_\Delta$ correlations for the soft component (center) and hard component (right) of the $y_t \otimes y_t$ correlations.\label{tat-fig2}}
\end{figure}

We have also measured p-p two-particle correlations~\cite{jeffp} on transverse momentum $p_t$ ($0.15 \leq p_t \leq 6.0$ GeV/c) transformed to transverse rapidity $y_t$. We observe large-scale correlation structures also consistent with p-p collisions having soft (string) and semi-hard (minijet) components. Fig.~\ref{tat-fig2} (left panel) shows correlation function $C(y_{t1},y_{t2}) = \rho_{sibling} - \rho_{mix}$ for STAR data. At low $y_t$ a soft component (string fragments) falls rapidly with increasing $y_t$. At higher $y_t$ the structure is centered around $y_{t1}=y_{t2}\simeq2.6$ ($p_{t1}=p_{t2}\simeq 1.0$~GeV/c) and attributed to correlated fragments from semi-hard parton scatters ({\em cf} distributions in Fig.~\ref{tat-fig1} (right panel)). 

These $y_t \otimes y_t$ correlations provide a cut space for analysis of axial momentum correlations~\cite{jeffp}. We construct joint autocorrelations on axial {\em difference variables} $\eta_{\Delta}\equiv \eta_{1}-\eta_{2}$ (pseudorapidity) and $\phi_{\Delta} \equiv \phi_{1}-\phi_{2}$ (azimuth angle) as pair-number ratio histograms. The right two panels in Fig.~\ref{tat-fig2} show autocorrelations from two $y_{t}$ selections in the left panel. The center panel (soft component) shows a broad peak on $\eta_\Delta$ caused by local charge conservation (charge ordering) on the fragmenting string. The depression on $\phi_\Delta$ near zero represents local transverse momentum conservation. The narrow peak is conversion electron pairs. The right panel (hard component) shows a near-side peak symmetric about $\eta_{\Delta}=\phi_{\Delta}=0$ and a broad $\eta_{\Delta}$-independent away-side ridge. The near-side peak represents  fragments distributed about a common jet thrust axis, and the away-side ridge correlates particles between opposing jets. Persistence of those structures to low hadron $p_{t}$ ($\sim 0.5$ GeV) may indicate that {\em minimum-bias} partons down to $p_t \sim 1$ GeV fragment to `minijets' of one or a few hadrons.

\section{$\langle p_t \rangle$ Fluctuations and $p_t$ Correlations in Au-Au}
 \label{mpt}

The scale (bin-size) dependence of event-wise mean transverse momentum $\langle p_{t} \rangle$ fluctuations~\cite{qingjun} can be inverted~\cite{invert} to obtain joint $p_t$ autocorrelations on pseudorapidity and azimuth angle difference variables representing velocity/temperature correlations. The structure and centrality dependence of those components suggest that the principal origin is minijets, altered by a dissipative color medium in the more central Au-Au collisions. 
\begin{figure}[h]
\vspace{-.0in}
\begin{tabular}{ccc}
\begin{minipage}{.31\linewidth}
\epsfysize .99\textwidth
\xfig{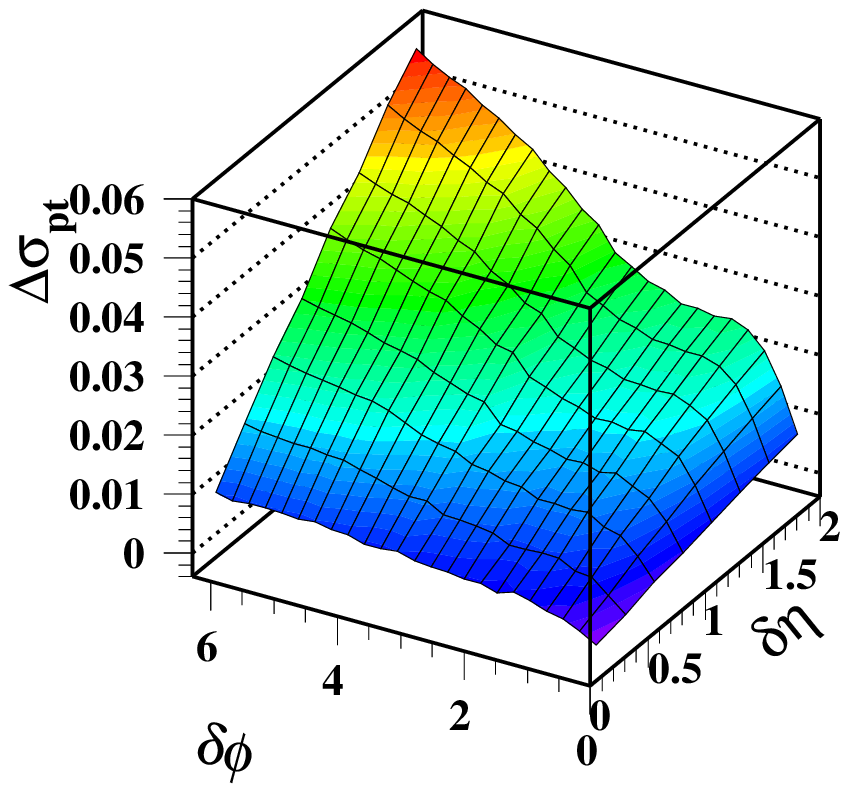}
\end{minipage} &
\begin{minipage}{.31\linewidth}
\epsfysize .99\textwidth
\xfig{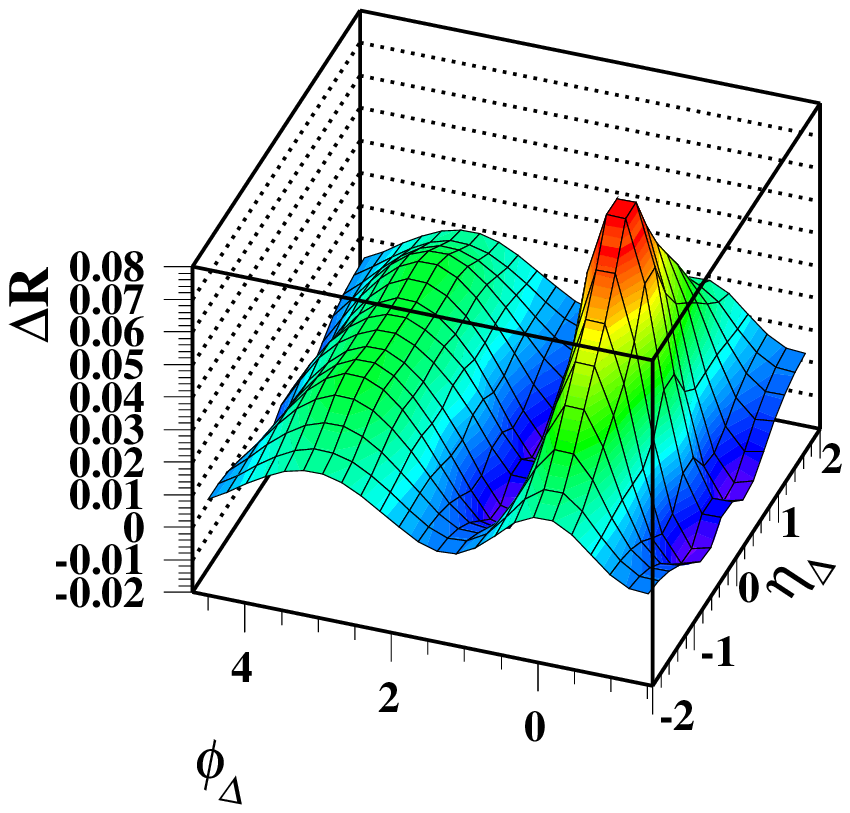}
\end{minipage} &
\begin{minipage}{.31\linewidth}
\epsfysize .75\textwidth
\xfig{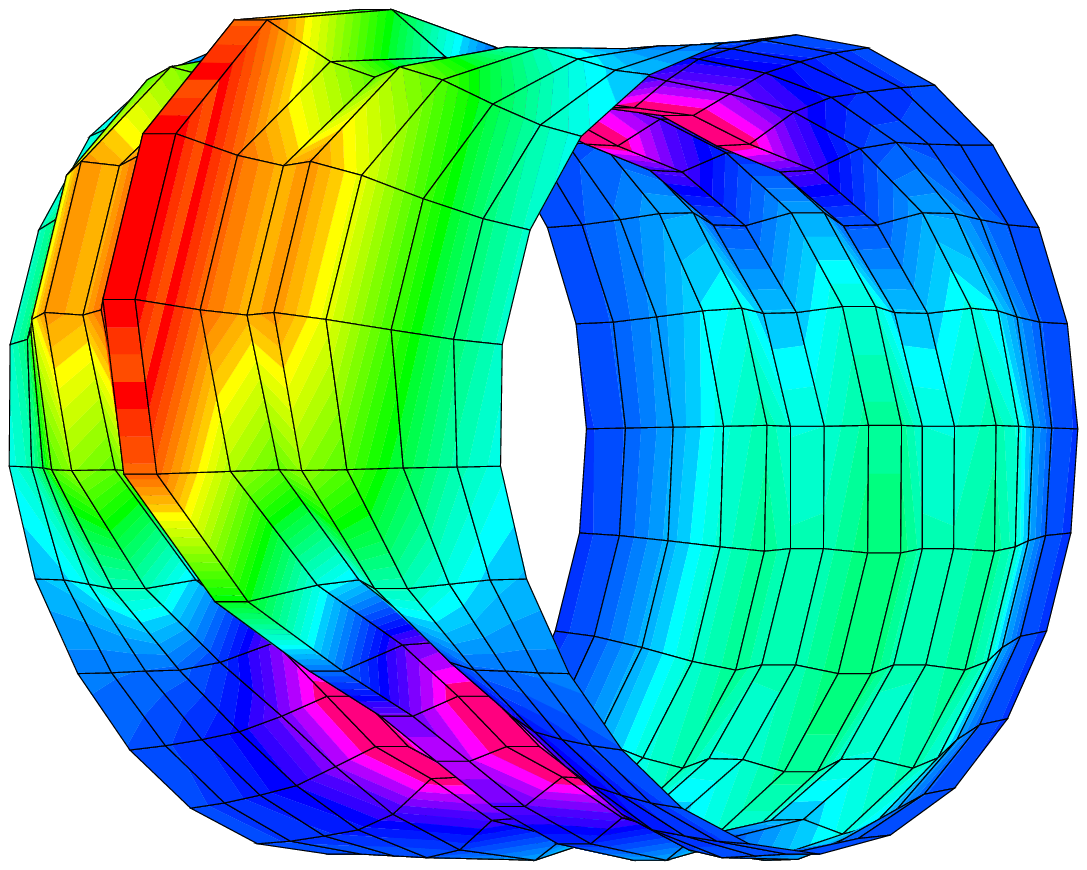}
\end{minipage}
\end{tabular}
\caption{Fluctuation scale dependence, inversion to $p_t$ autocorrelation and same autocorrelation distribution in cylindrical format.\label{tat-fig3}}
\end{figure}
Difference factor $\Delta \sigma_{p_t:n}$ is related to a {\em variance-difference} measure of nonstatistical $p_t$ fluctuations by $\Delta \sigma^2_{p_t:n} \equiv 2 \sigma_{\hat p_t} \Delta \sigma_{p_t:n}$. The variance difference is related to joint $p_t$ autocorrelation $\Delta R$ by integral equation
\bea \label{inverse}
\Delta \sigma^2_{p_t:n}(m \, \epsilon_\eta, n \, \epsilon_\phi) &\equiv& 
 4 \hat p_t^2\, \sum_{k,l=1}^{m,n} \epsilon_\eta \epsilon_\phi \, K_{mn;kl}  \,  \Delta R_{kl}(p_t:n;\epsilon_\eta,\epsilon_\phi).
\end{eqnarray}
In Fig.~\ref{tat-fig3} (left panel) is the variation with bin sizes ($\delta \eta,\delta \phi$) of $\langle p_t \rangle$ fluctuation measure $\Delta \sigma_{p_t:n}$ for central Au-Au collisions at $\sqrt{s_{NN}} = 200$ GeV up to the limiting STAR acceptance. The 2D joint autocorrelation on difference variables (center panel) obtained by inverting Eq.~(\ref{inverse}) has a near-side peak and away-side ridge which depend separately and strongly on collision centrality, and can be compared with Figs.~\ref{tat-fig2} (right panel) and \ref{tat-fig4} (left panels). The same distribution in cylindrical format (right panel) reveals `necking' on azimuth difference variable $\phi_\Delta$. This autocorrelation summarizes the complex velocity/temperature structure of multiple minijets and the bulk medium in each collision, averaged over many central Au-Au collisions.

\section{Parton Deformation in Au-Au Collisions}
  
\begin{figure}[h]
\vspace{-.0in}
\begin{tabular}{ccc}
\begin{minipage}{.3\linewidth}
\epsfysize 1.05\textwidth
\xfig{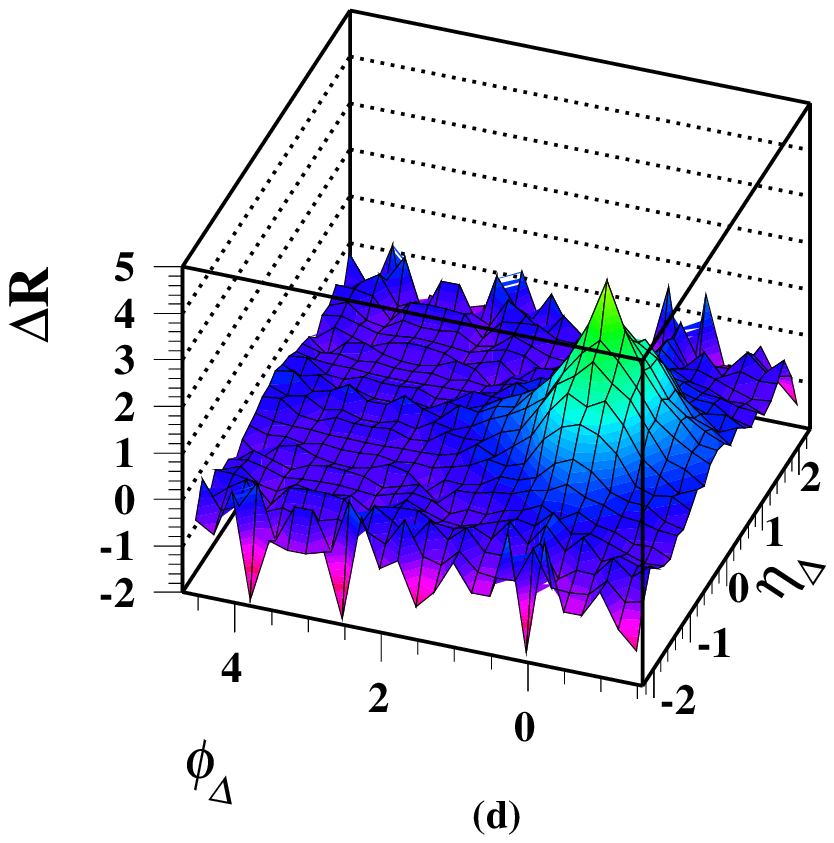}
\end{minipage} &
\begin{minipage}{.3\linewidth}
\epsfysize 1.05\textwidth
\xfig{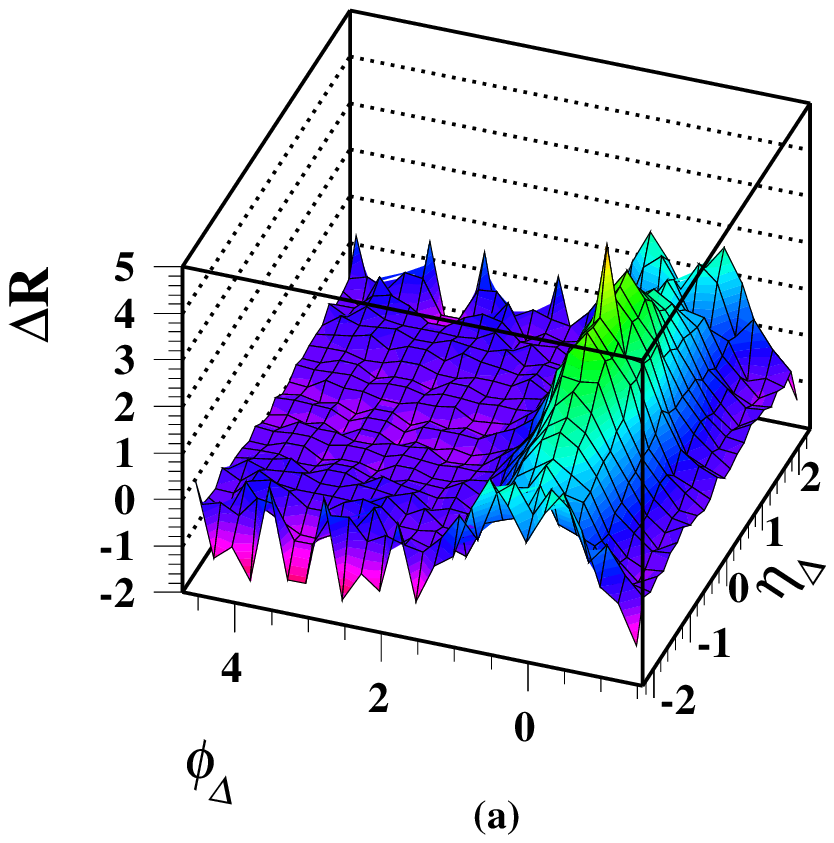}
\end{minipage} &
\begin{minipage}{.31\linewidth}
\epsfysize 1\textwidth
\xfig{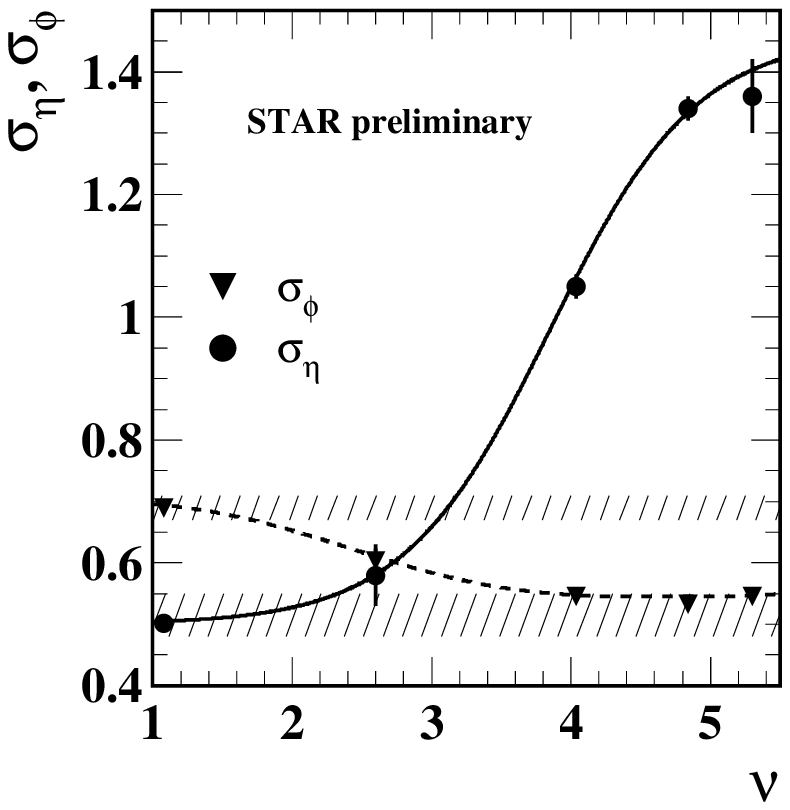}
\end{minipage}
\end{tabular}
\caption{Joint number autocorrelations with multipoles removed for peripheral (d) and central (a) events, and minijet width parameters $vs$ centrality $\nu$.\label{tat-fig4}}
\end{figure}

Charge-independent (CI) joint number autocorrelations on difference variables $\eta_\Delta$ and $\phi_\Delta$ were measured for particles with $0.15 \leq p_t \leq 2$~GeV/$c$ from Au-Au collisions at $\sqrt{s_{NN}} = 130$~GeV~\cite{aya1}. In Fig.~\ref{tat-fig4} (left panels) are autocorrelations for peripheral (d) and central (a) collisions with multipole components representing elliptic flow ($v_2$) and momentum conservation ($v_1$) removed, revealing minijet fragment angular distributions (conversion-electron pairs contribute to (0,0) bins). The ($\eta,\phi$) minijet angular correlations, symmetric about the origin in p-p and peripheral HI collisions (d), are strongly deformed in central collisions (a). Joint autocorrelations were fitted with a model function including azimuthal dipole and quadrupole terms and a 2D gaussian. Fitted gaussian widths $\sigma_\eta$ and $\sigma_\phi$ (right panel) $vs$ centrality measure $\nu$ (mean participant path length as number of encountered nucleons) show a dramatic centrality dependence.
Minijet peak structure varies in angular shape with centrality from a symmetric shape on ($\eta_\Delta,\phi_\Delta$) for peripheral collisions to a highly asymmetric peak for central collisions. This trend can be interpreted as a transition from {\em in vacuo} parton fragmentation similar to p-p collisions in peripheral heavy ion collisions to strong coupling of minimum-bias partons to a longitudinally-expanding color medium in central collisions. This result is inconsistent with pQCD-based models of jet quenching but may favor some recombination models.

\section{Centrality Evolution of Hadronization Geometry}
 
\begin{figure}[h]
\vspace{-.0in}
\begin{tabular}{ccc}
\begin{minipage}{.29\linewidth}
\epsfysize 1.05\textwidth
\xfig{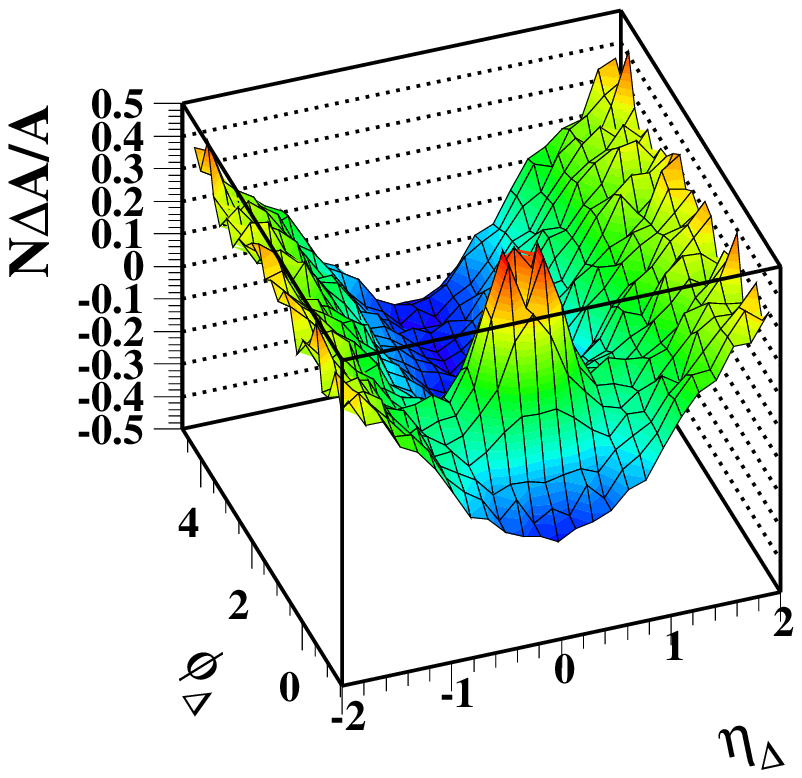}
\end{minipage} &
\begin{minipage}{.3\linewidth}
\epsfysize 1.1\textwidth
\xfig{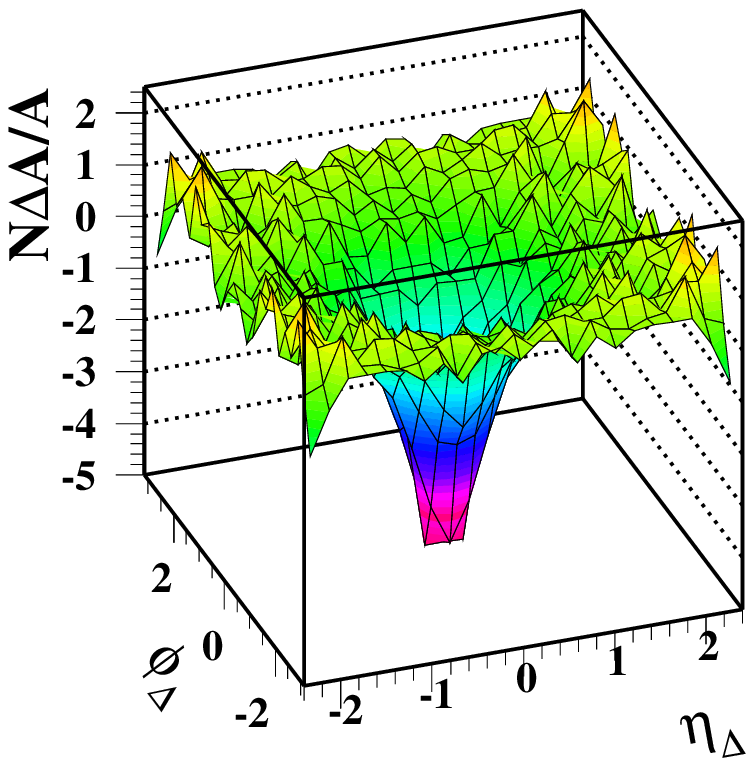}
\end{minipage} &
\begin{minipage}{.31\linewidth}
\epsfysize 1\textwidth
\xfig{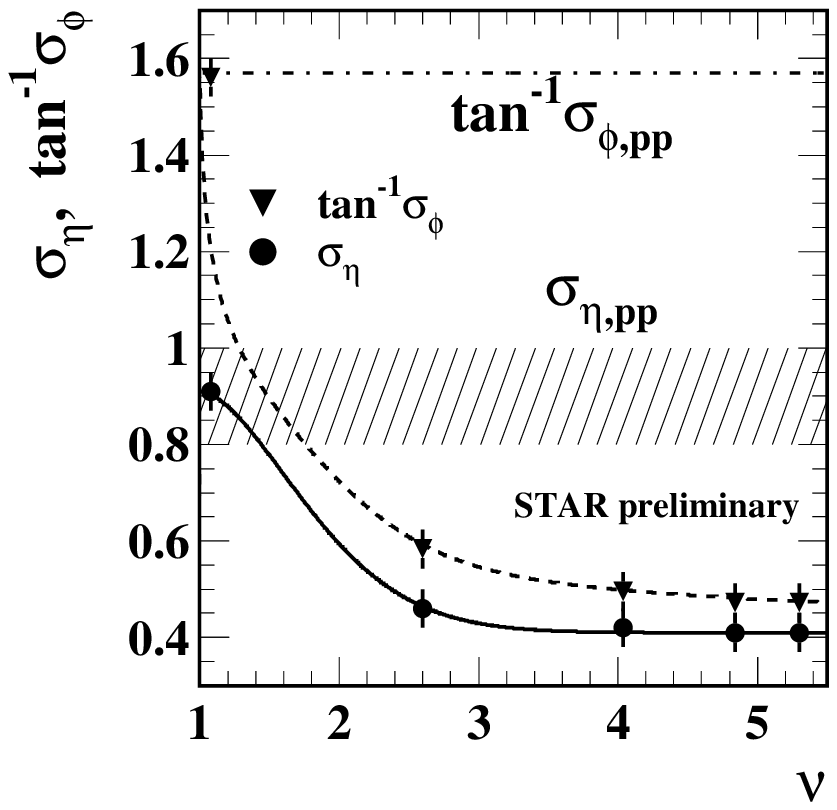}
\end{minipage}
\end{tabular}
\caption{Charge-dependent number autocorrelations for p-p collisions, central Au-Au collisions, and width parameters {\em vs} centrality measure $\nu$.\label{tat-fig5}}
\end{figure}

Charge-dependent (CD) joint number autocorrelations on difference variables $\eta_\Delta$  and $\phi_\Delta$ were measured for primary charged hadrons with $0.15 \leq p_t \leq 2$~GeV/$c$ from Au-Au collisions at $\sqrt{s_{NN}} = 130$~GeV~\cite{aya1}. Large-scale correlation structures, not predicted by theory, are consistent with a change in the geometry of hadron emission with increasing centrality of Au-Au collisions. In p-p collisions charge-dependent correlations are dominated by a 1D gaussian on $\eta_\Delta$ indicative of charge-ordering on the axial string as shown shown by 200 GeV p-p data in Fig.~\ref{tat-fig5} (left panel)~\cite{jeffp}. In central Au-Au collisions CD correlations are dominated by a peak at (0,0) nearly exponential in shape and nearly symmetric on ($\eta,\phi$). Fig.~\ref{tat-fig5} (center panel) shows CD correlations for the most central event class in this study: the 1D gaussian peak on $\eta_\Delta$ has disappeared and the amplitude of the central 2D exponential peak is eight times as large. Joint autocorrelations for four centrality classes were fitted with a model consisting of a 2D exponential with independent widths on $\eta_\Delta$ and $\phi_\Delta$, and a 1D gaussian on $\eta_\Delta$. Fitted width parameters are shown (right panel) {\em vs} centrality measure $\nu$.
These data are consistent with local charge conservation or canonical suppression of net charge fluctuations. The strong variation of correlation structure from p-p to peripheral Au-Au to central Au-Au can be interpreted as follows: Hadronization geometry evolves from 1D color-string fragmentation in p-p collisions to exponentially-attenuated 2D charge-ordered emission from a hadron-opaque medium in central Au-Au collisions. These results are qualitatively inconsistent with standard HI collision models and predictions of QGP-related strong suppression of net-charge fluctuations.
  
\section{Energy Dependence of $\langle p_t \rangle$ Fluctuations}
 
\begin{figure}[h]
\vspace{-.0in}
\begin{tabular}{ccc}
\begin{minipage}{.3\linewidth}
\epsfysize 1.05\textwidth
\xfig{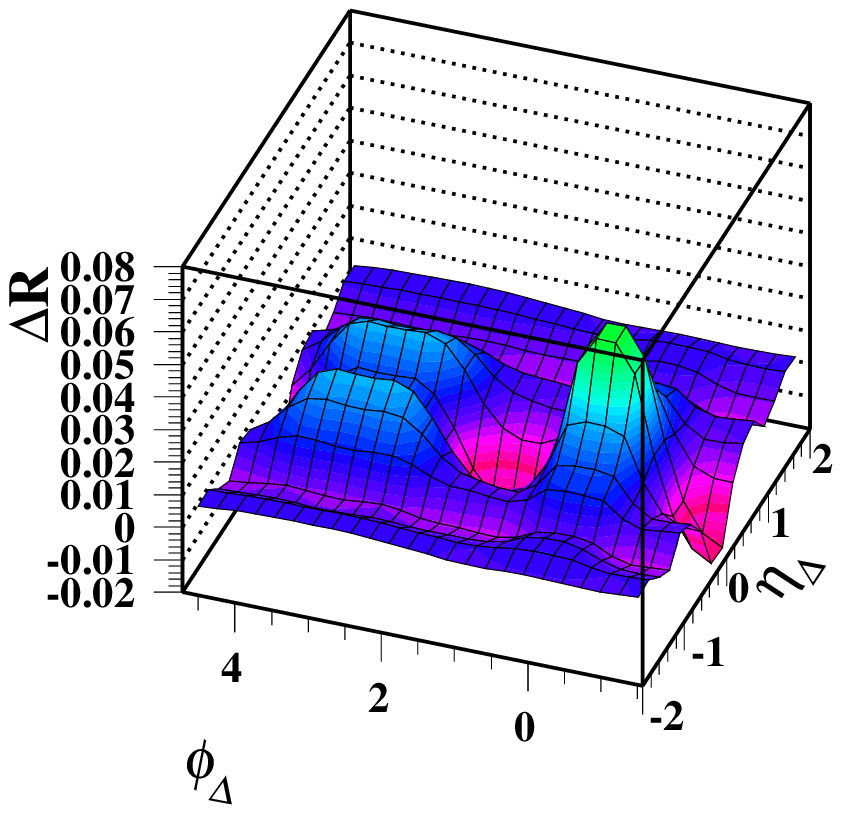}
\end{minipage} &
\begin{minipage}{.3\linewidth}
\epsfysize 1.05\textwidth
\xfig{trainor-3b.eps}
\end{minipage} &
\begin{minipage}{.31\linewidth}
\epsfysize 1\textwidth
\xfig{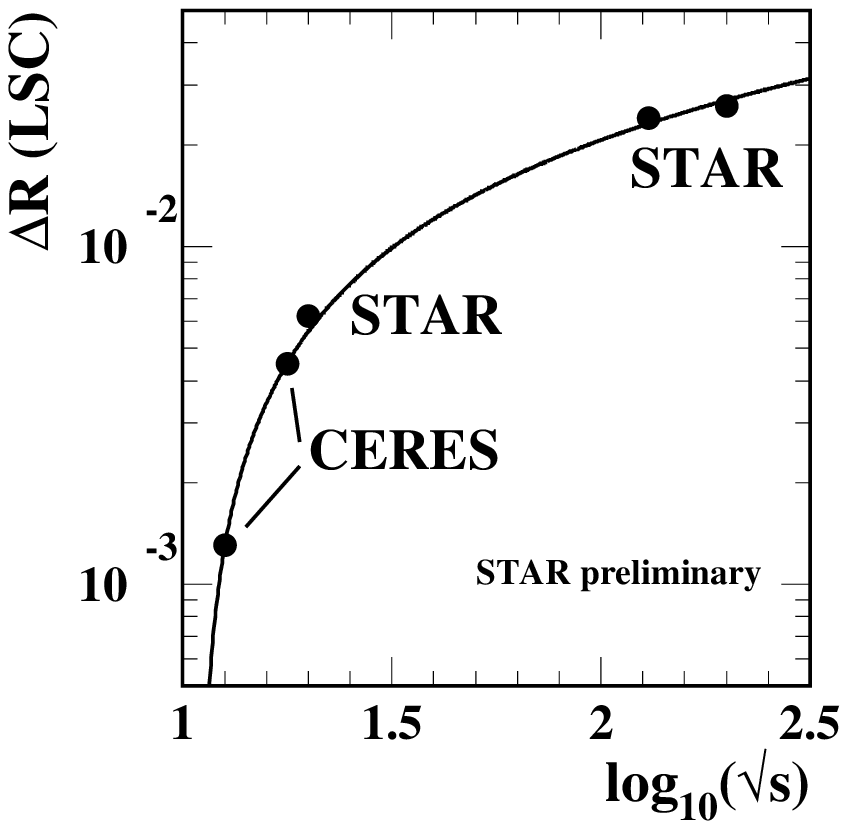}
\end{minipage}
\end{tabular}
\caption{STAR $p_t$ autocorrelations at 20 and 200 GeV, and energy dependence of large-scale correlations (LSC) from SPS to RHIC energies.\label{tat-fig6}}
\end{figure}

Controversy has emerged over the $\sqrt{s_{NN}}$ dependence of $\langle p_t \rangle$ fluctuations in heavy ion collisions from SPS to RHIC energies. Analysis of $p_t$ autocorrelations ({\em cf}\, Sec.~\ref{mpt}) at 20 and 200 GeV indicates a strong and monotonic increase with collision energy, with nontrivial $\langle p_t \rangle$ fluctuations first appearing at a threshold of observability near 10 GeV. Analysis based on `temperature fluctuations' on the other hand~\cite{ceres} suggests that $\langle p_t \rangle$ fluctuations are nearly independent of collision energy over this interval. In Fig.~\ref{tat-fig6} (left two panels) is a comparison of {\em per particle} $p_t$ autocorrelations inferred from STAR $\langle p_t \rangle$ fluctuation scale dependence at 20 and 200 GeV~\cite{qingjun}, with the same detector and analysis system in each case. A dramatic difference in fluctuations/correlations between SPS ($\sim 20$ GeV) and RHIC ($\sim 200$ GeV) energies is evident. The right-most panel shows the trend on collision energy (12.6, 17.8, 20 130 and 200 GeV) of the large-scale correlation (LSC) component of $p_t$ autocorrelations projected onto difference variable $\eta_\Delta$. This projection permits direct comparison with CERES' measurements of $\Phi_{p_t}(\delta \eta) \approx \Delta \sigma_{p_t:n}(\delta \eta)$ variation with pseudorapidity bin size~\cite{ceres}. LSC, defined as the autocorrelation amplitude at {\em large} $\eta_\Delta$, is distinguished from small-scale correlations (SSC), which especially at SPS energies are dominated by quantum (HBT) and Coulomb correlations. We observe that the principal mechanism for $\langle p_t \rangle$ fluctuations at RHIC is minijets. The energy dependence of the LSC component is therefore of considerable interest: what is the energy trend of semi-hard parton scattering, as manifested by correlated hadron fragments, down to the lowest relevant collision energy?  

\section{Conclusions}

By comparing correlation structures in p-p and Au-Au collisions obtained by several analysis methods we gain important new information about the changing geometry of hadronization and the interaction of low-$p_t$ or minimum-bias partons with the colored medium generated in the more central Au-Au collisions. We find that minimum-bias partons are sensitive to the velocity structure of the medium, and the hadronization process is dramatically different from that in elementary collisions.

\vfill\eject

\end{document}